\documentclass[onecolumn,showpacs,amsmath,amssymb,prd,nofootinbib]{revtex4}
\usepackage{graphicx}
\def\sss{\scriptscriptstyle}
\def\^#1{^{\sss #1}}
\def\_#1{_{\sss #1}}
\def\gmn{g_{\sss\mu \nu}}
\def\Gmn{g^{\sss\mu \nu}}

\def\Gab{g^{\sss\alpha\beta}}

\def\G#1{g^{\sss#1}}

\def\a{\alpha}
\def\b{\beta}
\def\c{\gamma}
\def\d{\delta}
\def\m{\mu}
\def\n{\nu}
\def\r{\rho}
\def\l{\lambda}

\def\der#1{{}_{\sss,#1}}
\def\deru#1{{}_{\sss,}{}^{\sss#1}}
\def\cd#1{{}_{\sss;#1}}

\def\ten#1#2{^{\sss#1}{}_{\sss#2}}
\def\tendu#1#2{_{\sss#1}{}^{\sss#2}}
\def\R#1#2{R^{\sss#1}{}_{\sss#2}}
\def\T#1#2{T^{\sss#1}{}_{\sss#2}}

\def\f#1{#1-form}

\def\div{\vec{\hbox{\mqq\char'162}}\cdot}
\def\grad{\vec{\hbox{\mqq\char'162}}}
\def\curl{\vec{\hbox{\mqq\char'162}}\times}
\def\div{\vec\nabla\cdot}
\def\grad{\vec\nabla}
\def\curl{\vec\nabla\times}
\def\sss{\scriptscriptstyle}
\def\^#1{^{\sss #1}}
\def\_#1{_{\sss #1}}
\def\beq{\begin{equation}}
\def\eeqno#1{\label{#1}\end{equation}}

\def\rarrow{\rightarrow }

\def\dleft{\rlap{{\it D}}\raise 8pt\hbox{$\scriptscriptstyle\Leftarrow$}}
\def\dright{\rlap{{\it
D}}\raise 8pt\hbox{$\scriptscriptstyle\Rightarrow$}}

\def\cmss{{\rm ~cm~s^{-2}}}

\def\az{a\_{0}}
\def\azs{a\_{0}\^2}

\def\l0{\ell\_{0}}

\def\rar{\rightarrow}
\def\s{\sigma}

\def\C{\Gamma}
\def\Cs{\bar\C}
\def\Up{\Upsilon}

\def\l{\lambda}

\def\f{\phi}

\def\r{\rho}

\def\m{\mu}
\def\n{\nu}

\def\G{\mathcal{G}}
\def\M{\mathcal{M}}

\def\F{\mathcal{F}}

\def\L{\mathcal{L}}

\def\d{\delta}

\def\a{\alpha}

\def\xlimin{{x\rarrow\infty \atop{\raise 1pt\hbox to 30pt{\rightarrowfill}}}}
\def\limlim#1#2{{#1\rarrow #2 \atop{\raise 1pt\hbox to 30pt{\rightarrowfill}}}}

\def\vq{{\bf q}}

\def\grad{\vec\nabla}
\def\div{\vec \nabla\cdot}
\def\gf{\grad\phi}

\def\lM{\ell\_M}
\def\baz{\bar a_0}
\def\R{\mathcal{R}}
\def\T{\mathcal{T}}
\def\Q{\mathcal{Q}}

\def\dsr{\ell\_{\Lambda}}

\def\vx{{\bf x}}

\def\hmn{h\_{\m\n}}
\def\emn{\eta\_{\m\n}}

\def\hab{h\_{\alpha\beta}}

\def\gh{g\^{1/2}}
\def\Emn{\eta\^{\mu \nu}}

\def\St{\mathcal{S}}
\def\Tmn{T_{\m\n}}

\def\hgmn{\hat g\_{\m\n}}
\begin{document}
\title{Noncovariance at low accelerations as a route to MOND}
\author{Mordehai Milgrom} \affiliation{Department of Particle Physics and Astrophysics, Weizmann Institute of
Science, Rehovot 76100, Israel}

\begin{abstract}
MOND has limelighted the fact that Newtonian dynamics and general relativity (GR) have not been verified to any accuracy at very low accelerations -- at or below the MOND acceleration $\az$: Without invoking made-to-measure ``dark matter'',  Newtonian dynamics (and hence general relativity) fail in accounting for galactic dynamics at such low accelerations. In particular, we do not have evidence that all the cherished, underlying principles of Newtonian dynamics or GR, such as locality or Lorentz invariance, still apply in the MOND limit. I discuss the possibility that the principle of general covariance might not apply in this limit. This would be in line with suggestions that general covariance, where it does hold, is only an emergent, hence approximate, property of relativistic dynamics. This idea also resonates well with MOND, which hinges on accelerations, for example because the existence of an effective absolute inertial frame is natural in MOND. Relaxing general covariance affords more freedom in constructing candidate MOND theories. For example, it may permit constructing pure-metric, local MOND theories, which is thought impossible with general covariance.
I exemplify this with a MOND-oriented, oversimplified, noncovariant theory, where the gravitational Lagrangian is $\L_M\propto \lM^{-2}\F(\lM\^{2}\R)$, where $\R= \Gmn (\C\ten{\c}{\m\n}\C\ten{\l}{\l\c}-\C\ten{\c}{\m\l}\C\ten{\l}{\n\c})/2$ is the (nonscalar) first-derivative part of the Ricci scalar $R$.  $\C\ten{\c}{\m\n}$ is the Levi-Civita connection of a metric, $\gmn$, which couples minimally to matter, and $\lM=c^2/\az$ is the MOND length, which is of cosmological magnitude, being, e.g., of the order of the de Sitter radius of our Universe.
This $\L_M$ gives a covariant theory in the high-acceleration limit by requiring that $\F(z)\rar z+\zeta$, for $z\gg 1$, which gives GR with a cosmological constant $\zeta c^{-4}\az^2$. In the MOND limit $\F'(z\ll 1)\propto z^{1/2}$.
In the nonrelativistic limit the metric has a solution of the form $\gmn\approx \emn-2\f\d\_{\m\n}$, as in GR, but the potential $\f$ solves a MOND, nonlinear Poisson analog.
This form of $\gmn$ also produces gravitational lensing as in GR, only with the MOND potential.
I show that this theory is a fixed-gauge expression of bimetric MOND (BIMOND), with the auxiliary metric constrained to be flat. The latter theory is thus a covariantized version of the former {\it \'{a}-la} St\"{u}ckelberg. This theory is also a special case of so-called $f(\Q)$ theories -- aquadratic generalizations of ``symmetric, teleparallel GR'', which are, in turn, also equivalent to constrained BIMOND-type theories.

\end{abstract}
%\pacs{04.50.-h  98.52.Eh  98.80.-k}
%\keywords{}
\maketitle

\section{introduction}
\label{introduction}
MOND\footnote{MOND originally stood for ``Modified Newtonian dynamics'', but has since attained a wider meaning, for example as a result of the advent of relativistic versions.} \cite{milgrom83a} is a paradigm of dynamics that arguably supplants Newtonian dynamics and general relativity (GR) in the realm of galaxies and the Universe at large, in a manner that obviates dark matter, and possibly dark energy. References \cite{fm12,milgrom14} are recent reviews of MOND.
\par
MOND departs greatly from standard dynamics at low accelerations, around and below some acceleration constant, $\az$, that MOND introduces. At high accelerations, $g\gg\az$, a MOND theory should tend to standard dynamics. In the opposite limit, $g\ll \az$, MOND dynamics become space-time scale invariant \cite{milgrom09}.
\par
MOND predicts a large number of ``galactic MOND laws'', most of which were listed already in the original MOND trilogy \cite{milgrom83a,milgrom83b,milgrom83c}, and they are discussed in detail in Refs. \cite{fm12,milgrom14a}. Some of these laws supersede Kepler's laws of planetary motions and other Newtonian relations, such as the virial theorem, in the low-acceleration regime. Some of them have no parallels in standard dynamics, as they pertain to the transition from high accelerations to the low ones. These laws constitute strict relations and strong correlations between the observed baryon distribution and dynamical properties of galaxies, which have been amply tested and vindicated.
\par
The constant $\az$ appears in many of these predictions and has been consistently determined to be $\az\approx 1.2 \times 10^{-8}\cmss$.
It has been noted since the advent of MOND (e.g., Refs. \cite{milgrom83a,milgrom89,milgrom99}) that $\az$ is near in value to some cosmologically significant accelerations:
\beq \baz \equiv 2\pi \az\approx a\_H(0)\equiv cH_0\approx a\_\Lambda\equiv c^2/\dsr, \eeqno{coinc}
where $a\_H\equiv cH$ is the acceleration associated with the cosmological expansion rate, $H$ (the Hubble-Lemaitre constant), and $a\_H(0)$ is its present-day value, and  $\dsr=(\Lambda/3)^{-1/2}$ is the radius associated with $\Lambda$ -- the observed equivalent of a cosmological constant. Defining the MOND length as
\beq \lM\equiv c^2/\az,  \eeqno{mondlength}
galactic dynamics and cosmology tell us that $\lM\sim \dsr\sim \ell\_H$, where $\ell\_H\equiv c/H_0$ is the Hubble distance today.
This numerical ``coincidence'', if fundamental, may have far-reaching ramifications for MOND, and for gravity in general (e.g., Ref. \cite{milgrom15}).
\par
Unless one invokes large quantities of ``dark matter'' in galactic systems, the predictions of Newtonian dynamics (hence of general relativity) greatly disagree with the observed dynamics of these systems.
The success of MOND phenomenology highlights the fact that these clashes occur, consistently, at low accelerations -- around and below $\az$. In the least, this is telling us that Newtonian dynamics and GR have not been tested and shown to hold in this MOND region of low accelerations.
\par
Known relativistic theories that incorporate MOND phenomenology have generally involved extensions of GR that depart from it at low accelerations. A question that arises when attempting such modifications is: `which of the basic principles that underlie GR -- such as locality, the weak equivalence principle, local Lorentz invariance, and general covariance -- are to be retained in such modified-dynamics theories?'
These principles of standard dynamics are supported by experiment and observation. But, they have not all been tested in the region of low accelerations that is of relevance to MOND; so they may be broken for $g\lesssim \az$, while well obeyed for $g\gg\az$. Relinquishing any of these principles may widen the scope of possible MOND extensions, may open the way for such extensions more easily, and may also help understand MOND's deeper origin.\footnote{We know that the weak equivalence principle -- the universality of free fall -- still holds to some accuracy at the low accelerations relevant for MOND. This is evidenced, for example, by the fact that different objects, such as stars of many types, gas clouds, and individual atoms, have consistent rotational speeds at the same orbital radius around the centers of disc galaxies, even at low accelerations.}
\par
If this is hard to palate, we have only to remember the long-cherished underlying principles of classical, Newtonian dynamics that have had to be abandoned in the quantum and relativistic regimes.
\par
Because MOND revolves around accelerations, in a MOND theory one has to identify system attributes with the dimensions of acceleration, which are to be compared with $\az$ as a criterion of whether we are in the high- or low-acceleration regime. This is similar, conceptually, to the need in the context of quantum mechanics, to define system parameters, $J$, with the dimensions of angular momentum (the dimensions of Planck's constant, $\hbar$) -- such as an action parameter or an angular momentum. $J/\hbar$ then appears in various expressions deduced from the theory, and for $J/\hbar\gg 1$ we are in the classical limit. Similarly, in the context of relativity, the theory compares with the speed of light parameters with the dimensions of velocity -- such as relative velocities or (square root of) gravitational potentials.
\par
GR is a local, generally covariant theory, derivable from an action, with a single metric as the gravitational degree of freedom in four-dimensional space-time. Its field equations are of second order, and special relativity is its ``no-gravity'' limit. Lovelock's theorem \cite{lovelock71} tells us that GR (with possibly a cosmological constant) is the only such theory (see also the discussion of general theoretical constraints on theories in Ref. \cite{clifton12}).

In addition, if we try to write a MOND-based theory with all these properties we encounter an obstacle: We cannot write a covariant expression for an acceleration involving first derivatives of the metric. For example, candidate acceleration parameters that can be constructed from a metric are the connections, but these are not tensors.\footnote{We can define tensor ``accelerations'' using higher derivatives of the metric, for example $c^2R^{1/2}$, with $R$ the Riemann curvature, which could underlie $\lM\^{-2}f(\lM\^2 R)$ theories. But these do not have the correct nonrelativistic (NR) MOND limit (see Ref. \cite{milgrom15a}).} If we try to modify the matter action, for example particle actions, then again we cannot construct from the world line an acceleration that is a diffeomorphism tensor. The covariant acceleration on a geodesic vanishes, and a quantity such as $d^2x\^\m/d\tau^2$, which does not necessarily vanish, is not a vector. A theory that makes use of such nontensor quantities will not be generally covariant.
\par
Most existing relativistic MOND theories circumvent the above obstacle by relinquishing, in the low-acceleration regime, one of the above-mentioned properties of GR. As emphasized above, this approach need not conflict with experimental or observational constraints.
\par
For example, scalar-tensor theories \cite{bm84}, TeVeS \cite{bekenstein04},  MOND-Einstein-ether theories \cite{zlosnik07}, bimetric MOND (BIMOND) \cite{milgrom09b}, and the bimetric massive gravity of Ref. \cite{blh17} involve gravitational degrees of freedom in addition to the metric. This enables one to define covariant (e.g. scalar) accelerations from first derivatives of these extra degrees of freedom. These extra degrees of freedom decouple, or are otherwise irrelevant, in the high-acceleration limit.
\par
The nonlocal theory of Refs. \cite{soussa03,deffayet14} circumvents the obstacle by relinquishing locality, which is restored in the high-acceleration limit.
\par
The MOND theories of Refs. \cite{bm11}, and \cite{sanders11} (inspired by Ho\v{r}ava-Lifshitz-type modifications) break local Lorentz invariance in the MOND regime in a way that allows one to define accelerations from first derivatives.\footnote{These theories can equivalently be described by Lorentz-invariant theories with an added degree of freedom, the khronon field, from which accelerations can be built.}
\par
But, it is also possible that nature defines an absolute inertial frame -- a fact that can be felt only at low accelerations. This would make it possible to define absolute accelerations.
As was discussed in Ref. \cite{milgrom99} the quantum vacuum defines such a frame: Observers in a laboratory that accelerate with respect to the vacuum can measure this acceleration using the Unruh effect and the way it varies across the laboratory (see more on this in Sec. \ref{mondvac}).

\par
The MOND theories proposed to date -- relativistic and nonrelativistic -- are very useful in various regards: Some provide crucial proofs of concept, for example that one can construct relativistic MOND theories with correct gravitational lensing. Some may point in interesting directions where to look for the origin of MOND phenology. Some provide tools with which to perform calculations within MOND -- such as galaxy formation and interactions -- ensuring from the outset that the basic MOND tenets are satisfied, and the salient MOND predictions are inherited, while at the same time the standard requirements such as the weak equivalence principle and the usual conservation laws are guaranteed (e.g. Refs. \cite{tiret08,candlish15,thomas17,bilek17}).
\par
It has to be said, however, that we do not yet have a fully satisfactory, all-encompassing theory to underlie the MOND paradigm. Not only are there still important aspects of the mass anomalies that are not yet addressed by MOND (such as cosmology and structure formation), but the theories we now have seem to have the hallmarks of approximate, effective descriptions, which emerge from a more fundamental theory, which we still lack (see Sec. \ref{mondngc}).
\par
Yet another way to circumvent the above barrier is to forgo general covariance (aka diffeomorphism invariance) in the low-acceleration regime.
My main purpose here is to discuss different aspects of this option, and see if it might conduce to better understanding of MOND's origin.
\par
Indeed, the possibility has been raised and discussed that general covariance (as well as other gauge symmetries) might not be fundamental (e.g., \cite{seiberg06,anber10,sindoni11,ho13,witten18}), to wit, that general covariance is not enjoyed by the yet-to-be-found underlying theory for GR -- which might not look anything like GR, having perhaps a completely different arena, degrees of freedom, etc. Instead, general covariance  emerges only under some limiting circumstances of the more fundamental theory, where GR is a good description. Because it is in the quantum-gravitational regime that GR's shortcomings are an accepted fact, these suggestions, naturally, have in mind an underlying theory (without general covariance) that would account for quantum gravity. Thus, general covariance is thought to emerge in this context at low energies (or for distances much larger than the planck length).
\par
But the main idea underlying MOND is that GR is also wanting in another regime of phenomena -- that of low accelerations. And so we may borrow the idea of emergent general covariance, and suggest that the fundamental theory that may underlies MOND is not generally covariant, and that general covariance emerges as an attribute of the less fundamental, approximate theory that we now call MOND. But it does so only in the high-acceleration regime $g\gg cH_0$, or $g\gg c^2/\dsr$.
This ties well with the idea that MOND is emergent, as discussed further in Sec. \ref{mondngc}.
\par
Note that general covariance is not a symmetry of GR that would be lost in the deep-MOND limit -- it does not relate different physical configurations of a given system, as symmetries of physical theories do -- it is rather invariance to reparametrization, reflecting redundancy in the way we describe a physical system, and relating different descriptions of exactly the same physical state of the system. In contradistinction, the high- and low-acceleration regimes in MOND do differ in their symmetries, the latter enjoying scale invariance, unlike the former.
\par
There are strong experimental and observational limits on a certain class of noncovariant departures from GR \cite{anber10}. But these were all obtained in very-high-acceleration systems, such as the inner Solar System and do not exclude departures in the deep-MOND limit (see Sec. \ref{ncv}).
\par
Extensions of GR with a preferred inertial frame have a log history (e.g., Refs. \cite{rosen74,katz85}). But in the context of MOND this possibility is, arguably, better motivated and germane.
\par
One should note the following general caveat: As a result of relation (\ref{coinc}), a system, say of mass $M$ and size $R$,
is both relativistic ($MG/R\sim c^2$), and in the deep-MOND limit ($MG/R^2\lesssim\az$), only if $R\gtrsim\lM$, namely if the system is of cosmological dimensions. This may tell us that it does not make sense to look for a relativistic MOND theory that is not part and parcel of cosmology.
\par
My main aim here is to point out these facts concerning departures from general covariance in the deep-MOND limit, and that, as already stated, forgoing this principle (a) is not in conflict with measurements, and (b) may open a promising route to the fundamentals of MOND.
\par
In Sec. \ref{mondngc}, I enlarge on the naturalness of noncovariance in MOND, and in Sec. \ref{ncv}, I discuss a simple example of a noncovariant, modified-gravity, relativistic MOND theory.

\section{Noncovariance and MOND\label{mondngc}}
As discussed above,
MOND points to a ``hiding place'' for noncovariance in the domain of small accelerations, where general covariance has not been tested to my knowledge.
But beyond that, general covariance breakdown, while not required, may be natural in MOND. There are several hints for this, which I now describe.
\subsection{MOND as we know it is arguably emergent}
The point has been repeatedly pressed that MOND as we now know it must be an effective, approximate formulation that emerges from a deeper theory, which might look very different from any theory we now use, and may be underlain by new principles, not enjoyed by existing theories  (e.g., \cite{milgrom02,milgrom15}).
\par
One strong indication for this is the ``coincidence'' (\ref{coinc}): If seemingly unrelated constants that appear in a theory are (numerically) simply related, it is a hint that one of them might be calculable from the others in a way that is understood only in a more fundamental, underlying theory. Such is the case for the constants appearing in thermodynamics (of ideal gases): the gas constant, Avogadro's number, and Boltzmann's constant, hinting that thermodynamics emerges from statistical mechanics. Another example is the approximate, constant-gravitational-acceleration theory for near-Earth-surface phenomena, where the Galilei acceleration $g$, the Earth radius, $R\_\oplus$, and the escape speed from the surface $V_{es}$, are related by $g=V^2_{es}/2R\_\oplus$, a relation that can be understood only within Newton's universal gravity theory.
\par
In the case of MOND, it is not yet clear which of the constants involved, $c$, $\az$, and $\lM$, are the more fundamental and which are derived. For example, arguments along those given in Ref. \cite{milgrom99} (see Sec. \ref{mondvac} below) suggest that, as in the above example of the Galilei acceleration,
the approximately de Sitter geometry of the Universe characterized by a radius $\sim\lM$, enters and ushers $\az$ into local dynamics.
Or, take the picture described in Ref. \cite{milgrom19} whereby local MOND dynamics emerge in a brane-world picture. There, $\az$ appears as the acceleration on masses living on the brane universe, due to a force field acting in the embedding space of the brane. But this same force is responsible for balancing the brane itself at radius $\lM$ against its own tension. This dual role of the external acceleration is what gives rise to relation (\ref{coinc}); so $\lM$ and $\az$ emerge together.
\par
Another indication for MOND, as now formulated, being emergent is that all action-based MOND theories propounded to date involve some ``interpolation function'' between the high- and low-acceleration regimes that is put in by hand -- a function of one variable of the form $A/\az$, where $A$ is some characteristic system acceleration. This is a hallmark of effective field theories.
\par
We can take a lesson from quantum theory, where, indeed, various such interpolating functions appear in quantum expressions describing different phenomena. These are functions of a variable of the form $J/\hbar$, where $J$ is some attribute with the dimensions of $\hbar$. Examples are the blackbody function, which interpolates between the low-frequency, Rayleigh-Jeans regime and the high-frequency, quantum regime, or the expression for the specific heat of solids, interpolating between the high-temperature, classical, Dulong-Petit expression and the quantum, low-temperature behavior. However, these expressions are not the underlying theory itself. While all involve an interpolation between the classical ($\hbar\rar 0$) and the quantum ($\hbar\rar \infty$) limits, they differ from phenomenon to phenomenon, and, of course, they do not appear as fundamental functions in the underlying quantum theory itself. It is likewise expected that there is a fundamental origin theory for MOND in which such interpolating functions are not introduced at the fundamental level but that emerge in different contexts, under the circumstances to which we apply MOND today.
\par
In recognition of this status of MOND, there are, indeed, many attempts to derive MOND phenomenology from some more fundamental starting points (e.g., Refs. \cite{blt08,pikhitsa10,ho10,kt11,lc11,kk11,pa12,bk16,smolin17,verlinde17}, and see Refs. \cite{milgrom14,milgrom15} for more references).
These attempts have, however, not yet lead to a full-fledged underlying theory for MOND.
\subsection{MOND and the quantum vacuum as an absolute, global inertial frame \label{mondvac}}
It has been suggested \cite{milgrom99}, in connection with MOND, that the quantum vacuum may define an absolute inertial frame. This is based on the observation that an observer  can measure its absolute acceleration with respect to the vacuum by using the Unruh effect.\footnote{It was also explained how a finite-size, rigid observer can measure both the magnitude and the direction of its acceleration -- the latter is along the gradient of the Unruh temperature across the observer.} It has been proposed that the interaction with the vacuum endows bodies with inertia -- an occurrence that is rife in other contexts of physics (such as electrons in solids or, indeed, mass renormalization in quantum field theories). It was also suggested that in a flat Minkowski background, standard inertia emerges. But it is not so in the nearly de Sitter background we live in (as evinced by the dominant contribution to the energy density in the Universe from a cosmological constant).
\par
It was shown (heuristically) how in a de Sitter background we may get inertia that behaves according to the basic MOND tenets, with, furthermore, $a\_\Lambda$ playing the role of MOND's $\az$. This comes about because an accelerated observer in a de Sitter background sees an Unruh effect that depends both on the observer's motion with respect to the vacuum and on the global de Sitter radius.\footnote{An inertial observer then sees itself in a bath of black body with the Gibbons-Hawking temperature of the de Sitter.}
\par
It was argued in Ref. \cite{milgrom99} that the different dynamical behaviors of bodies with accelerations with respect to the vacuum comes about as follows: Observers with $a\gg a\_\Lambda$ probe, using their Unruh wavelength $\l\_U\equiv c^2/a$, distances that are $\ll\dsr$, and so behave as in a flat background, while for observers with $a\gtrsim a\_\Lambda$, the Unruh wavelength does feel the de Sitter nature of the background.\footnote{Several later works have built on similar ideas, e.g., Refs. \cite{pikhitsa10,ho10,kk11,pa12}.}
\par
The heuristic arguments of Ref. \cite{milgrom99} were based on what we know about the Unruh effect for the ideal case of a constant-acceleration observer in an exact de Sitter background.
It was also argued there that since the Unruh effect is not local in time,\footnote{What an observer measures at a given time depends on the whole of its (past) motion.} the emergent laws of dynamics (e.g., inertia) are also generically time nonlocal. For unknown reasons, in the high-acceleration regime, $a\gg a\_\Lambda$, the emergent dynamics become local, but perhaps not so in the MOND regime. {\it There is no evidence from galaxy phenomenology that dynamics there is time local},\footnote{Much of the most reliable data come from disc-galaxy rotation curves. There, the orbits are circular and the instantaneous represents well the long-term behavior. In other stationary systems, such as virialized isolated galaxies, this is true on average for the system. For nonstationary systems -- such as small satellites in orbit around mother galaxies -- it may well be, but hard to pinpoint that their present equilibrium dynamics is determined also by time-nonlocal, memory effects.} and such nonlocality would even tie in well with nonlocality in MOND \cite{milgrom94,soussa03,deffayet14}.
\par
As discussed above, the quantum vacuum as an absolute inertial frame breaks general covariance just as the cosmic microwave background defines a preferred rest frame, and breaks Lorentz invariance. In the latter case, we know the (electromagnetic) effects of the cosmic microwave background on bodies and they do not go far in dynamically breaking Lorentz invariance beyond some matter-of-principle aspects. However, the unknown microscopic effects of the vacuum -- as reflected, e.g., in the puzzles related to the cosmological constant -- could be important enough to affect the required dynamics in a way that MOND phenomenology emerges (see e.g., Ref. \cite{klinkvol19} and references therein for a possible microscopic description of the quantum vacuum).
\par
The quantum vacuum is Lorentz invariance, so such a picture does not introduce a preferred Lorentz frame. It is also highly isotropic; so it is not expected to break experimental limits on the anisotropy of inertia.
\par
If all this is correct, we would need to understand why the emergent dynamics, which are presumably nonlocal in general, do become local to a very high accuracy in the high-acceleration regime,
and also, more pertinent to our discussion here, why the dynamics become generally covariant to high accuracy in this high-acceleration regime. Does this too result from this limit corresponding formally to a Minkowski vacuum as opposed to a de Sitter one?
To boot, we will need to understand why in the opposite, very-low-acceleration limit, the dynamics become scale invariant, at least in the weak-field limit (in the relativistic sense, e.g., when $|\gmn-\emn|\ll 1$).
\subsection{The MOND external-field effect}
More generally, MOND might require the measurability of an absolute acceleration.
One manifestation of this that is quite amenable to observational tests is the so-called external-field effect (EFE) of MOND. It has been noticed from the advent of MOND (see, e.g., Refs. \cite{milgrom83a,bm84}) that generically, in MOND, the internal dynamics of a small subsystem falling freely in the gravitational field of a mother system is affected by the presence of the external field. Reference \cite{milgrom14a} discusses at length when and how such an effect follows from the basic assumptions of MOND. But the fact is that this effect is present in all MOND theories proposed to date.
\par
Some observations pointing to the EFE in action in galactic systems are described, e.g., in Refs. \cite{mm13,wk15,haghi16,hees16,mcgaugh16a,banik18}
\par
One dramatic manifestation of the effect is that an external acceleration $\gg\az$ annihilates MOND effects within a system, even if the internal accelerations are $\ll\az$ \cite{milgrom83a}.
In other words, standard dynamics is restored within a system subject to a high external acceleration field.
This explains, for example, why MOND effects can hardly be detected with Earth-bound experiments, where the gravitational acceleration is $\sim 10^{11}\az$, and kinematic acceleration due to Earth rotation and revolution are $\sim 10^8\az$.
\par
An implication of the EFE in MOND is that the strong equivalence principle does not hold. This in itself does not imply breakdown of general covariance, as clearly there are covariant theories -- in particular, covariant MOND theories -- that break the strong equivalence principle.
\par
The presence of the EFE also does not necessarily imply the existence of a global, inertial frame, like the one the vacuum could define, as discussed in Sec. \ref{mondvac}, or like the one discussed in the next section. For example, in BIMOND, such a frame is defined locally by the auxiliary metric.
\par
However, the opposite might be true, namely that noncovariance manifests itself in the EFE, making the physics inside a system dependent on the system's acceleration with respect to the inertial frame.

\section{An example of modified gravity \label{ncv}}
In GR both the gravitational (Einstein-Hilbert) action and the matter action are generally covariant. These imply, correspondingly, the Bianchi identities for the Einstein tensor, and the divergencelessness of the matter energy-momentum tensor, implying, in turn, conservation of energy and momentum.
\par
If general covariance emerges for the high-acceleration regime, but does not hold in the deep-MOND regime, we can expect that this applies not only to the gravitational action (as in ``modified gravity''), but to the matter actions as well. Ignoring this might lead to inconsistencies.
\par
Nonetheless, since my aim here is not to explore thoroughly theories and their implications, only  to point to noncovariance as a possible route to MOND, I demonstrate my point with
an (arguably oversimplified) modifications of only the gravity sector.
\par
As explained in Ref. \cite{milgrom15a}, the straightforward extensions of GR that replace the Einstein-Hilbert Lagrangian density, $R$, by some (dimensionless) function of it $\ell^{-2}f(\ell\^2 R)$ ($\ell$ is some length) -- so-called $f(R)$ theories -- cannot give MOND phenomenology (except, perhaps, in a radically modified form as suggested, e.g., in Refs. \cite{bernal11,barrientos17}).
This is because if -- as in GR -- we can write the metric in the NR limit as
$\gmn=\emn-2\f\d\_{\m\n}$, then, to leading order in $\f/c^2$, we have $R\propto c^{-2}\Delta\f$, where $\f$ is the NR potential. But this cannot lead to MOND phenomenology, where it is required that the NR Lagrangian be of the form $f(\lM c^{-2}|\gf|)=f(|\gf|/\az)$.
\par
However, relinquishing general covariance in the MOND regime does allow us to construct a simple $f(R)$-like action with MOND phenomenology, while still retaining the metric as the only gravitational degree of freedom.
These theories  look superficially like $f(R)$ theories but they are quite different (see more on this below). In such theories one defines the gravitational Lagrangian density as a function of polynomial expressions built from only the affine connections $\C\ten{\l}{\m\n}$ with all indices contracted, such as $\Gmn\C\ten{\l}{\m\a}\C\ten{\a}{\l\n}$. In particular, since MOND provides us with a ready length scale, $\lM$, we can construct such dimensionless expressions from the dimensionless $\lM\C\ten{\l}{\m\n}$, with Lagrangian densities of the form $\lM^{-2}\F(\lM\C\ten{\l}{\m\n})$.
Such expressions are not coordinate scalars; hence the breakdown of general covariance. Importantly, these Lagrangians contain only first derivatives of the metric. Also, they are still invariant to linear coordinate transformations, which do not affect $\C\ten{\l}{\m\n}$, and so they are, in particular, Lorentz invariant. The second property would be consistent, for example, with the idea that it is the quantum vacuum that defines an inertial frame, the vacuum being Lorentz invariant.

\par
Start with the Einstein-Hilbert covariant action underlying GR (without a cosmological-constant term):
\beq I=-\frac{c^4}{16\pi G}\int\gh R~d^4x
 +I\_M(\gmn,\psi_i),  \eeqno{gedat}
where $I\_M$ is the matter action where the metric couples standardly to matter degrees of freedom, $\psi$ (the determinant of the metric is $-g$.)
The Ricci scalar
\beq R=\Gmn R\ten{\a}{\m\a\n}=\Gmn (\C\ten{\a}{\m\a}\der{\n}-\C\ten{\a}{\m\n}\der{\a})-2\R, \eeqno{ricci}
where
\beq \R= \Gmn\R_{\m\n}; ~~\R_{\m\n}\equiv \frac{1}{2}(\C\ten{\c}{\m\n}\C\ten{\l}{\l\c}-\C\ten{\c}{\m\l}\C\ten{\l}{\n\c}). \eeqno{nirta}

\par
As is well known (e.g. Ref. \cite{schrodinger50}),
\beq \gh R=2\gh\R+ q\^\m\der{\m};\eeqno{arar}
so $R$ in eq. (\ref{gedat}) may be replaced with $2\R$ without changing the content of the theory, since this changes the Lagrangian by a complete divergence. However, since $\R$ is not a diffeomorphism scalar, the resulting action is not manifestly covariant.
\par
In the context of MOND, I here propound a generalization of GR in which the gravitational Lagrangian, $R$, in the Einstein-Hilbert action is replaced by
\beq \L_M=2\lM\^{-2}\F({\lM\^{2}\R}). \eeqno{molag}
[The normalization of $\R$ is chosen so that the argument of $\F$ becomes $(\gf/\az)^2$ in the NR limit , where $\f$ is the nonrelativistic MOND potential; see Sec. \ref{nonrel}].
This gives a theory that is no longer covariant because the change in $\R$ under diffeomorphisms appears in the argument of $\F$ and cannot be ignored as a complete divergence.
\par
If, for example, as discussed in Sec. \ref{mondvac}, the quantum vacuum defines a dynamically effective inertial frame, the connections appearing in the Lagrangian (\ref{molag}) may be taken, as those defined there. Any local inertial Lorentz frame will do, as $\L$ is a Lorentz scalar, and the vacuum is locally Lorentz invariant.\footnote{Strictly speaking, globally, for a de Sitter vacuum, we should rather speak of the de Sitter isometry group, which contracts to the Poincar\`{e} group locally.}
\par
Henceforth we take $c=1$, unless stated otherwise; so $\lM=\az\^{-1}$). Thus,
the MOND action is
\beq I=-\frac{\azs}{8\pi G}\int\gh \F(\R/\azs)~d^4x
 +I\_M(\gmn,\psi_i).  \eeqno{gedatas}
\par
According to the basic tenets of MOND, general relativity (possibly with a cosmological constant) should be restored in the limit $\lM\rar\infty$ ($\az\rar 0$). We thus require  \beq\F(z)\rar \F(\infty)+z~~~{\rm for}~~~ z\rar\infty, \eeqno{limama}
 where $\F(\infty)$ is a dimensionless constant. So, in this limit
\beq I\rar-\frac{1}{16\pi G}\int\gh [2\R+2\azs\F(\infty)]~d^4x
 +I\_M(\gmn,\psi_i).  \eeqno{gedatut}
This is the action of general relativity with a cosmological constant $\Lambda=-\azs\F(\infty)$. The resulting cosmological constant is naturally of the order of $\az\^2$ [if $\F(\infty)=O(1)$], as observed and described in relation (\ref{coinc}).
\par
In the opposite limit, defined by $\lM\rar 0$ with $G/\lM$ fixed, scale invariance of the nonrelativistic, deep-MOND limit dictates (see Sec. \ref{nonrel})
 \beq \F(z)\rar \F(0)+\a z\^{3/2}~~{\rm for}~z\rar 0. \eeqno{limumu}
  For the standard normalization of $\az$, $\a=2/3$.
Generally, for $\R\not \gg \lM^{-2}$ noncovariance remains.
\par
In light of the coincidence (\ref{coinc}), the transition occurs at the cosmologically significant value of $\R\approx \R_\Lambda= \dsr^{-2}$. But note that since (for nonrelativistic systems) $\R\sim (\gf)^2$, this can occur in cosmologically small systems where $|\gf|\sim\az$.
\par
Reference \cite{anber10} discusses experimental and observational constraints on a certain class of departures from general covariance. Their non-generally-covariant Lagrangian densities are, like our $\R$, index-contracted, quadratic expressions in $\C\ten{\a}{\m\n}$, but which, unlike our $\R$, cannot be written as a diffeomorphism scalar up to a total derivative. Unlike MOND, such theories do not involve a new dimensioned constant, under which manifestations of noncovariance can be hidden. \footnote{In principle, there is a large family of such noncovariance Lagrangians that one can write as there are many contracted polynomial expressions, and one can take various functions of them. Reference \cite{anber10} gives some examples of  Lagrangians quadratic in $\C\ten{\l}{\m\n}$ (so there is no need to use a length constant) as benchmarks for testing general-covariance breakdown.}

\subsection{Relation to BIMOND \label{bimond}}
In the above discussion, I arrived at the action (\ref{gedatas}) starting from Schr\"{o}dinger's rendition of the Einstein-Hilbert Lagrangian $\R$, and used the fact that MOND provides us a natural scale length $\lM$, to construct the Lagrangian as a function of $\R$.
\par
We can also arrive at this Lagrangian through another interesting route, starting from the bimetric formulation of relativistic MOND (BIMOND) \cite{milgrom09b}.
\par
BIMOND  is a class of (covariant) theories where the gravitational sector involves two metrics: one, $\gmn$, to which matter couples (minimally) and a ``twin'' metric $\hgmn$.
Different versions introduce different degrees of symmetry between the two metrics (e.g. with respect to whether there is a twin matter sector, raising and lowering indices, and definition of the volume forms).
Here, we need $\hgmn$ as only auxiliary, so I consider the version where only $\gmn$ is used for all raising and lowering of indices, and in the volume element of the interaction term.
\par
The BIMOND action is then taken as
\beq  I=-\frac{1}{16\pi G}\int[ \gh R +{\hat g}\^{1/2}\hat R
 -2\gh\lM\^{-2}\M]d\^4x
 +I\_M(\gmn,\psi\_i),  \eeqno{biact}
where $\M$ is a dimensionless scalar function of quadratic (dimensionless) scalars constructed from $\lM C\ten{\a}{\b\c}$, with the tensor (being the difference between two affine connections)
\beq C\ten{\a}{\b\c}\equiv\C\ten{\a}{\b\c}-\hat\C\ten{\a}{\b\c}.  \eeqno{condif}
Here, $\C\ten{\a}{\b\c}$ and $\hat\C\ten{\a}{\b\c}$ are the Levi-Civita connections of the two metrics, and $R$ and $\hat R$ are the Ricci scalars constructed from them.
$I\_M$ is the matter action with $\psi$ representing matter degrees of freedom.
\par
The quadratic scalars are of the form
\beq U\ten{\b\c\m\n}{\a\l}(\gmn,\hgmn)C\ten{\a}{\b\c}C\ten{\l}{\m\n},  \eeqno{juoit}
where in the more general version, the ``contraction factors'' $U$ are products of the metrics and their inverses. If we use only $\gmn$ in $U$, as here, there are five independent such quadratic scalars. However, in Ref. \cite{milgrom09b} and subsequent works on BIMOND, it was suggested to concentrate on one such scalar as an argument for $\M$ because of some special properties it has. This scalar argument is $-\lM^2\Up/2$, where
\beq \Up=\Gmn\Up\_{\m\n},~~\Up\_{\m\n}=
C\ten{\c}{\m\l}C\ten{\l}{\n\c} -C\ten{\c}{\m\n}C\ten{\l}{\l\c}.   \eeqno{rtudes}
To make contact with the action (\ref{gedatas}), we add to this BIMOND version a Lagrange-multipliers constraint that forces the auxiliary metric to be flat, by adding to the Lagrangian density a term\footnote{In Rosen's bimetric gravity \cite{rosen74}, the auxiliary metric is also forced to be flat
in this way. Unlike ours, his Lagrangian is quadratic in $C\ten{\a}{\b\c}$ but differs from the one that would give GR.}
\beq \l\tendu{\a}{\b\c\d}\hat R\ten{\a}{\b\c\d}. \eeqno{lagmul}
Varying the action over $\hgmn$ would give equations that determine the Lagrange multipliers, which we do not care about. Varying over the multipliers gives $\hat R\ten{\a}{\b\c\d}\equiv 0$, which means that there are coordinate choices $\xi\^\m$ for which $\hgmn$ is constant; so, there $\hat\C\ten{\a}{\b\c}\equiv 0$.
In such a frame (free up to linear coordinate transformations), $\Up=-2\R$ of our theory, and so this constrained BIMOND action, in the chosen frame, coincides with our action (\ref{gedatas}), with the identification $\F(z)=-\M(z)+z$.
\par
In any other frame, $x\^\m(\xi\^\s)$, we have for the auxiliary connection
\beq \hat\C\ten{\a}{\m\n}=\frac{\partial x\^\a}{\partial \xi\^\s}\frac{\partial^2 \xi\^\s}{\partial x\^\m\partial x\^\n}.   \eeqno{parata}
So in the fully covariant theory (i.e., without fixing the gauge to $\xi\^\m$) we have four gravitational degrees of freedom, $x\^\m(\xi\^\s)$, in addition to $\gmn$.
\par
The constrained version of BIMOND can thus be viewed as a covariantized  extension of our theory, {\it \'{a}-la} St\"{u}ckelberg, with $x\^\m(\xi\^\s)$ the added St\"{u}ckelberg fields.
\subsection{Other routes -- $f(\Q)$ theories}
There are additional routes to the action (\ref{gedatas}) as generalizations of GR.
For example, Refs. \cite{jimenez18,jimenez19} have recently discussed a theory governed by a Lagrangian density of the form (\ref{molag}), though not with MOND in mind, so not for the form of $\F$ and the value of $\lM$ relevant for MOND.
\par
They start from the so-termed symmetric, teleparallel GR (e.g., \cite{nester99,adak06}). This formulation of GR employs as (independent) gravitational degrees of freedom a metric $\gmn$ and an affine connection $\Cs\ten{\a}{\m\n}$ that is not the metric-compatible, Levi-Civita one for $\gmn$ (which we call $\C\ten{\a}{\m\n}$).\footnote{Namely, the covariant derivatives of $\gmn$ with respect to $\Cs\ten{\a}{\m\n}$ do not vanish, as they do with respect to $\C\ten{\a}{\m\n}$.}
Two requirements imposed on the connection are (a) $\Cs\ten{\a}{\m\n}=\Cs\ten{\a}{\n\m}$ is symmetric (can be imposed with Lagrange-multipliers term) and (b) it gives an identically vanishing curvature; so, absolute parallelism holds for this connection (hence the epithet ``teleparallel''). This is done, e.g., by adding a Lagrange-multipliers term like (\ref{lagmul}) in the Lagrangian. As an affine connection, $\Cs\ten{\a}{\m\n}$ transforms under diffeomorphisms like $\C\ten{\a}{\m\n}$
so
\beq L\ten{\a}{\m\n}\equiv\C\ten{\a}{\m\n}-\Cs\ten{\a}{\m\n} \eeqno{miute}
is a (symmetric) tensor, like the $C\ten{\a}{\m\n}$ in BIMOND.\footnote{It can be expressed in terms of the nonvanishing covariant derivatives of $\gmn$, with respect to $\Cs\ten{\a}{\m\n}$.}
\par
To formulate an equivalent of GR, the Lagrangian density is taken as the scalar quadratic in $L\ten{\a}{\m\n}$ with the same index combination as in $\R$ in eq. (\ref{nirta}), and in $\Up$ in BIMOND, namely,
\beq \Q= \Gmn\Q_{\m\n}; ~~ \Q_{\m\n}\equiv \frac{1}{2}(L\ten{\c}{\m\n}L\ten{\l}{\l\c}-L\ten{\c}{\m\l}L\ten{\l}{\n\c}). \eeqno{aserty}
It turns out that, for $\Cs\ten{\a}{\m\n}$ whose curvature vanishes, $\Q$ differs from the
Ricci scalar of $\gmn$, $R(\gmn)$, by a divergence. Hence, taking $\Q$ as the Lagrangian density gives an action that is equivalent to the Einstein-Hilbert one -- yielding an alternative formulation of GR.
\par
This is also essentially the construction used previously by Ref. \cite{katz85} to write a Lagrangian for GR with only first derivatives of the metric. It is a bimetric theory with the auxiliary metric taken as flat; so $\Cs\ten{\a}{\m\n}$, which is constrained to be flat can be thought of as the Levi-Civita connection of the flat metric of Ref. \cite{katz85}.
\par
As in all the theories we discuss here, it is possible to generalize also by considering other quadratic scalars, e.g., of the $C\ten{\a}{\m\n}$ in BIMOND or of $L\ten{\a}{\m\n}$ in symmetric, teleparallel theories. In fact, Rosen's bimetric theory \cite{rosen74} is a predecessor to the latter -- again, if we think of $\Cs\ten{\a}{\m\n}$
as the Levi-Civita connection of Rosen's flat, auxiliary metric.\footnote{Rosen wanted a theory that differs from GR; so his Lagrangian is not $\Q$. But he still chose it according to certain criteria, in particular, so that the nonrelativistic potential defining $g_{00}$ satisfies the Poisson equation with matter density as the source, so that Newtonian gravity obtains for slowly moving masses.}
\par
References \cite{jimenez18,jimenez19} then suggested generalizations of this theory, having in mind not MOND but intending to modify GR at high energies, introducing some relevant length scale $\ell$, and constructing Lagrangians that are more general functions $f(\ell^2\Q)$.
\par
Similar to what happens in the constrained version of BIMOND discussed above, for solutions of these theories, because $\bar R\ten{\a}{\b\c\d}[\Cs\ten{\l}{\m\n}]=0$, one can choose a coordinate frame in which $\Cs\ten{\l}{\m\n}\equiv 0$.
In this frame $L\ten{\l}{\m\n}=-\C\ten{\l}{\m\n}$, and so $\Q=\R$, and with the choice $\ell=\lM$, and $f=\F$, appropriate for MOND, we get our action (\ref{gedatas}).
\par
Such $f(\Q)$ theories are, likewise, equivalent to the constrained BIMOND-type theory described above:
With the imposition of teleparallelism and symmetry, the affine connection $\Cs\ten{\l}{\m\n}$ is, perforce, the Levi-Civita connection of some flat metric (which can be taken to have a Minkowskian signature) $\c\_{\m\n}$. So $L\ten{\a}{\m\n}=\C(g)\ten{\a}{\m\n}-\C(\c)\ten{\a}{\m\n}$ is just the $C\ten{\a}{\m\n}$ of BIMOND, with the flatness constraint on $\c\_{\m\n}$.
\par
Thus, the $f(\Q)$ theory can also be considered a covariantization of the theory governed by the action (\ref{gedatas}) {\it \'{a}-la} St\"{u}ckelberg.
\par
There
are treatments of cosmological models within $f(\Q)$ (e.g., \cite{jimenez19a,dialektopoulos19,lazkoz19}) where it was found that there are some advantages to adopting the specific form of $\Q$ above.  In the context of MOND, I do not see much point in applying this theory to cosmology, as in itself the theory still has some gaps (e.g., as regards the matter action), and I bring it here only as a heuristic example. Besides, MOND as we now know it and as described, for example, by the theory discussed here, is arguably only an ``effective field theory'' which is probably not a good approximation for treating cosmology.
\subsection{Field equation}
To get the field equation for the metric, we need to vary the action over $\Gmn$. Here I use a shortcut, taking advantage of our theory being a constrained version of BIMOND, as discussed in Sec. \ref{bimond}.
I thus take the BIMOND field equation derived in Ref. \cite{milgrom09b} for the version with only $\gmn$ used for contraction, and for the mixed-term volume element. In this substitute  $\hat \C\ten{\l}{\m\n}=0$ to get
  \beq \G_{\m\n}\equiv -2\F'\R\_{\m\n}
 +(\F'\St\ten{\l}{\m\n})\cd{\l}+
 \azs\F\gmn=
 8\pi G \Tmn,  \eeqno{misha}
where $\F=\F(z)$, with $z=\R/\azs$, and where\footnote{Note that $\F'$ is not a scalar, and  $\St\ten{\l}{\m\n}$ is not a tensor. The covariant derivative of $\F'\St\ten{\l}{\m\n}$ is understood as the standard expression in terms of the derivatives and the connection.}
\beq \St\ten{\l}{\m\n}\equiv\C\ten{\l}{\m\n}-\frac{1}{2}\d\ten{\l}{\m}\C\_{\n}-\frac{1}{2}\d\ten{\l}{\n}\C\_{\m}
 +\frac{1}{2}\gmn(\C\^{\l}- \C\^{*\l}), \eeqno{gyrtaz}
\beq \C\_{\n}\equiv \C\ten{\l}{\l\n},~~~
 \C\^{\l}\equiv g\^{\a\l}\C\_{\a},~~~
 \C\^{*\l}\equiv\Gab \C\ten{\l}{\a\b}, \eeqno{dunet}
and  $\Tmn$ is the matter energy-momentum
tensor.
\par
For $\F(z)=z=\R/\azs$, the action (\ref{gedatut}) reduces to that of GR, for which $\G_{\m\n}=-G_{\m\n}$, with $G_{\m\n}$ the Einstein tensor. From this we deduce the identity
\beq 2\R\_{\m\n}-\gmn \R
 -\St\ten{\l}{\m\n}\cd{\l}=G_{\m\n}. \eeqno{idad}
 So we can write the field equation as
  \beq \F'G_{\m\n}-\F'\der{\l}\St\ten{\l}{\m\n}
 -\azs(\F-z\F')\gmn= -8\pi G \Tmn.  \eeqno{mishadab}
\par
These equations  hold in our supposed preferred inertial frame (e.g., that of the quantum vacuum, if indeed it serves as such a frame). The covariantized versions using constrained BIMOND or $f(\Q)$ can be used if we want to describe the system in other frames.

\subsection{Weak-field limit}
I next consider the weak-field and the nonrelativistic limits, as they are more relevant to MOND phenomenology regarding galactic dynamics, and gravitational lensing.
\par
In the total absence of gravity -- no matter present, and no cosmological constant, which we neglect when discussing cosmologically small systems -- we expect (or require) that the geometry is flat. The field equation (\ref{mishadab}) then tells us that in our preferred frame
$\C\ten{\a}{\b\c}\equiv 0$ is a solution, which we take. The metric is thus a constant, since its derivatives are linear in $\C\ten{\a}{\b\c}$. We can use the invariance we still have to general linear coordinate transformation to choose coordinates where $\gmn=\emn$.
\par
The weak-field limit (WFL) applies for small departures from Minkowski:
\beq \gmn=\emn+\hmn,~~\hmn\ll 1,  \eeqno{gusta}
where only the lowest order in $\hmn$ is kept.
\par
The WFL of $\R$, $\bar\R(\hmn)$, is of the order of $(h\_{\m\n,\l})^2$. It can be seen that in $\F$ and its volume prefactor we can replace everywhere $\gmn$ by $\emn$, and $g\_{\m\n,\s}$ by  $h\_{\m\n,\s}$.
Then, $\C\^\a\_{\b\c}$ is linear in  $\hmn\der{\a}$, and
$\F$ becomes a function of variables of the form $(\hmn\der{\a}/\az)^2$. These variables are of zeroth order in our approximation, and so are all appearances of $\F$ in the WFL.\footnote{Despite the appearance of $h^2$ in them, $(\hmn\der{\a}/\az)^2$ are of zeroth order. The WFL corresponds to $\hmn\ll 1$, but $\hmn\der{\a}/\az$ are not necessarily small. In fact, they are very large in the Newtonian limit of MOND.}
[$\hmn\der{\a}$ are first order; $\az$ is first order; $\hmn\der{\a}/\az$ is zeroth order; $(\hmn\der{\a})^2$ is neglected relative to $\hmn\der{\a}\der{\b}$. So, e.g., $\R_{\m\n}$ and $\azs\F$ are neglected relative to the other term in the equation of motion. Only terms linear in $\C$ or $\az$ are kept.]
\par
Also,
\beq I\_M\approx \frac{1}{2}\int \hmn\T^{\m\n}d^4x, \eeqno{matu}
where $\T^{\m\n}$ is the Minkowskian energy-momentum tensors.
\par
The weak-field action is
\beq I\approx -\frac{\az^2}{8\pi G}\int \F(\bar\R/\azs)~d^4x+\frac{1}{2}\int \hmn\T^{\m\n}d^4x,  \eeqno{gepada}
where
\beq \bar\R=\frac{1}{8}[{{h\^{\n\r}}\_{,}}\^\c(h\_{\n\r,\c}-2h\_{\n\c,\r})
-{{h}\_{,}}\^\c(h\_{,\c}-2h\^{\r}\_{\c,\r})],  \eeqno{kiolio}
 is the weak-field form of $\R$.
 (Indices are raised and lowered with $\emn$ in the WFL.)

\par
The field equation then reads
\beq [\F'(\bar\R/\azs)\bar \St\ten{\l}{\m\n}]\_{,\l}
=8\pi G\T_{\m\n},  \eeqno{biupo}
where $\bar \St\ten{\l}{\m\n}$ is the WFL of the expression in eq. (\ref{gyrtaz}),
with
\beq \C\^\a\_{\b\c}\approx
\frac{1}{2}\eta\^{\a\s}(h\_{\b\s,\c}
+h\_{\c\s,\b}-h\_{\b\c,\s}).\eeqno{lalama}
Thus,
\beq \C\_{\n}=h\der{\n}/2,~~\C\^{\m}= \Emn \C\_{\n},~~\C\^{*\m}=h\^{\m\a}\der{\a}-(1/2)h\deru{\m},~~h=h\ten{\a}{\a}. \eeqno{gamudi}
\par
$\bar \St\ten{\l}{\m\n,\l}=
-\bar G\_{\m\n}(\hab)$ is (minus) the WFL of the Einstein tensor; it satisfies the Bianchi identity $\bar \St\ten{\l}{\m\n,\l,}{\^\n}=0$.

\subsection{Nonrelativistic limit\label{nonrel}}
In the NR limit (weak field and slow motions of the sources), the only nonvanishing component of $T_{\m\n}$ is $T_{00}=\r$. Also, the metric is time independent in this approximation.

\par
Write, then, the metrics, generally, as slightly perturbed from Minkowski:
 \beq \gmn=\emn-2\f\d\_{\m\n}+h\_{\m\n}, \eeqno{rutza}
Define $\f\equiv (\eta\_{00}-g\_{00})/2$; so $h\_{00}\equiv 0$.
Also, write the mixed metric elements as
$h\_{0i}=h\_{i 0}=q\_i$ (Roman letters are used for space indices).
\par
We then write the field equations to first order in the potentials
$\f$ and $h\_{\m\n}$ (noting that there is no time dependence):
\beq [\F'(\bar\R/\azs)\bar \St\ten{i}{\m\n}]\_{,i}
=8\pi G\r\d\_{\m 0}\d\_{\n 0}.  \eeqno{biuer}
\par
In our approximation,\footnote{Remember that because the metric derivatives,
connections, and curvature components are already first order, all
the metrics that are used to contract them can be taken as $\emn$.}
 $$ \C\ten{0}{00}=0,~~\C\ten{i}{00}=\C\ten{0}{0i}=\C\ten{0}{i0}
 =-{1\over 2}g\_{00,i}=\f\_{,i},~~\C\ten{i}{0j}=(q\_{i,j}-q\_{j,i})/2,~~
\C\ten{0}{ij}=-(q\_{i,j}+q\_{j,i})/2,$$
  \beq  \C\ten{i}{jk}=
 {1\over 2}(g\_{ij,k}+g\_{ik,j}-g\_{jk,i})=
 {1\over 2}(h\_{ij,k}+h\_{ik,j}-h\_{jk,i})
 +\f\_{,i}\d\_{jk}-\f\_{,j}\d\_{ik}-\f\_{,k}\d\_{ij}.
  \eeqno{litara}

 $$ \bar\St\ten{i}{00}=2\f\_{,i}+{1\over 2}(h\ten{j}{i}\der{j}-\bar h\_{,i}),~~\bar\St\ten{i}{0j}=\frac{1}{2}(q\_{i,j}-q\_{j,i})$$
 \beq
 \bar\St\ten{i}{jk}={1\over 2}(h\_{ij,k}+h\_{ik,j}-h\_{jk,i})
+{1\over
4}[2(\bar h\_{,i}-h\ten{m}{i}\der{m})\d\_{jk}-\bar h\_{,k}\d\_{ij}
-\bar h\_{,j}\d\_{ik}].
 \eeqno{gedtup}
Also
\beq \bar\R=(\gf)^2+\frac{1}{8}[{{h\^{ij}}\_{,}}\^k(h\_{ij,k}-2h\_{ik,j})
-{{\bar h}\_{,}}\^k(\bar h\_{,k}-2h\^{j}\_{k,j})]-\frac{1}{8}(q\_{i,j}-q\_{j,i})(q\_{i,j}-q\_{j,i}),  \eeqno{kiolar}
where $\bar h$ is the 3-D trace $\bar h=h\^i\_i$. Thus, $\vq$ enter the argument of $\F'$ as $(\curl\vq)^2$.
\par
The $(0j)$ components of eq. (\ref{biuer}) are
\beq [\F'(\bar\R/\azs)(q\_{i,j}-q\_{j,i})]\der{i}=0,\eeqno{paret}
which can be written as
\beq \curl[\F'(|\curl\vq|^2)\curl\vq]=0.  \eeqno{cular}
(There is also dependence of $\F'$ on $\vx$ through $\f$ and $h\_{ij}$.)
With the boundary condition that $\vq$ vanishes fast enough at $\infty$, this implies\footnote{The scalar product of $\vq$ with eq. (\ref{cular}) gives $\div(\vq\cdot\curl\vq \F')-(\curl\vq)^2\F'=0$.
So, integrating over space, with the boundary condition at $\infty$ gives $\int (\curl\vq)^2 \F'[(\curl\vq)^2]= 0$. But $\F'$ is non-negative, and vanishes only for $\bar\R=0$ (in the MOND version). Hence the non-negative integrand must vanish everywhere. This ignores the unlikely possibility that there is some finite volume where $\curl\vq\not=0$, but $\bar R=0$} $\curl\vq=0$. Thus, $\vq$ disappears from the other equations, since it appears as $\curl\vq$. Also, we can write $\vq=\grad u(\vx)$ for some function $u$. The field equations for this general static space-time (where the metric does not depend on the time coordinate) are invariant under a local change of the zero of time: $t=t'-v(\vx')$, $\vx=\vx'$, which changes $\vq$ by $\grad v(\vx)$, and does not change other elements of the metric. We can choose $v$ such that $\vq=0$.
\par
Look now at the six $(jk)$ equations
\beq [\F'(\bar\R/\azs)\bar \St\ten{i}{jk}]\_{,i}=0. \eeqno{iiijjj}
As was the case in BIMOND, the choice of the specific quadratic argument $\R$ results in
$\bar \St\ten{i}{jk}$ depending (linearly) only on $h\_{ij}$, and not on $\f$, as shown in eq. (\ref{gedtup}).
This means that $h\_{ij}=0$ is a solution if we impose that $h\_{ij}\rar 0$
fast enough at infinity. This is not the case for general choices of the quadratic argument of $\F$, since then $\f$ too appears in $\bar \St\ten{i}{jk}$.
\par
For the solution $h\_{jk}=0$, we have $\bar\R=(\gf)^2$, and $\bar\St\ten{i}{00}=2\f\_{,i}$; so
the $(00)$ equation is:
\beq \div[\F'(|\gf|\^2/\azs)\gf]
=4\pi G\r. \eeqno{raqua}
This is the nonlinear Poisson MOND theory of Ref. \cite{bm84} (dubbed AQUAL -- for aquadratic Lagrangian).
\par
We thus conclude that a NR system of masses has a gravitational field
 \beq \gmn=\emn-2\f\d\_{\m\n}, \eeqno{rutbit}
which has the same form as in GR and is characterized by one NR potential, $\f$, which, however, is not determined from the Poisson equation as in GR, but from the MOND equation (\ref{raqua}).
This means, for example, that, when interpreted by GR, light and massive, slow test bodies see the same potential.
\par
This result exemplifies the usefulness of the particular choice of argument of $\F$ we took in eq. (\ref{nirta}). With most other choices, $h\_{ij}=0$ is not a solution because $\bar\St\ten{i}{jk}$ depend also on $\f$ and do not vanish for $h\_{ij}=0$. In such a case, the NR metric is described by many potentials, and does not have the GR form. This has observationally unwanted implications for lensing, for example.\footnote{In this regard, the scalar $\R$ is not unique; see
footnote 14 in Ref. \cite{milgrom09b}.}
\subsection{Discussion}
\label{discussion}
Beyond phenomenology,
there are several matter-of-principle issues that need to be checked, before we can consider the above theory a serious candidate. I only list these issues here as they are beyond the present scope. These issues pertain also to the constrained BIMOND theories, and to $f(\Q)$ theories.
\par
In Ref. \cite{milgrom09b}, I showed that BIMOND enjoys a tractable Cauchy problem, but so far I was not able to show the same for the present theory.
Also, as already alluded to, questions related to modifying the matter actions remain open.
Another question is that of stability and the appearance of ghosts. And, not least, gravitational-wave propagation is now an important constraint in light of the observations of the recent neutron-star merger.

I thank Nathalie Deruelle for helpful comments and suggestions.

\clearpage
\end{document}